\documentclass[conference]{IEEEtran}
\usepackage{cite}

\ifCLASSINFOpdf
\usepackage[pdftex]{graphicx}
\DeclareGraphicsExtensions{.pdf,.jpeg,.png}
\else
\usepackage[dvips]{graphicx}
\DeclareGraphicsExtensions{.eps}
\fi
\usepackage{amsmath} 
\usepackage{amsfonts}
\usepackage{amsthm}

\usepackage{algorithmic}
\usepackage{algorithm}

\usepackage{array}
\usepackage{hhline}

\usepackage{subfig}

\theoremstyle{plain} 
\newtheorem{cachingState}{Proposition}
\newtheorem{spaceStairs}[cachingState]{Proposition}

\usepackage{times}

\begin{document}
%
\title{QoE-Aware Proactive Caching of Scalable Videos Over Small Cell Networks}

\author{\IEEEauthorblockN{Zhen Tong,
		Yuedong Xu,
		Tao Yang, 
		Bo Hu}
	\IEEEauthorblockA{Department of Electronic Engineering, Fudan University\\
		Email: \{14210720043, ydxu, taoyang, bohu\}@fudan.edu.cn}}


\maketitle

\begin{abstract}
The explosion of mobile video traffic imposes tremendous challenges on present cellular networks. 
To alleviate the pressure on backhaul links and to enhance the quality of experience (QoE) of video streaming service, small cell base stations (SBS) with caching ability are introduced to assist the content delivery. 
In this paper, we present the first study on the optimal caching strategy of scalable video coding (SVC) streaming in small cell networks with the consideration of channel diversity and video scalability.
We formulate an integer programming problem to maximize the average SVC QoE under the constraint of cache size at each SBS. 
By establishing connections between QoE and caching state of each video, 
we simplify the proactive caching of SVC as a multiple-choice knapsack problem (MCKP), and propose a low-complexity algorithm using dynamic programming. 
Our proactive caching strategy reveals the structural properties of cache allocation to each video based on their popularity profiles.
Simulation results manifest that the SBSs with caching ability can greatly improve the average QoE of SVC streaming service, 
and that our proposed caching strategy acquires significant performance gain compared with other conventional caching policies.
\end{abstract}


%

\section{Introduction}

Mobile video service is witnessing a tremendous growth nowadays. As forecasted by Cisco, 
mobile video traffic will increase 11-folds between 2015 and 2020, accounting for 75\% of 
total mobile data traffic by 2020 \cite{cisco_prediction}. 
To deal with this phenomenon, 
the industry is advocating the deployment of small cell base stations (SBS) to enable higher density spatial reuse of radio resources. 
However, a drawback of this approach is the huge expenditure to connect all the SBSs to the core network with fast backhaul links\cite{femtocell_survey}.
Meanwhile, video contents requested by users exhibit significant similarities, 
thus causing a large amount of redundant traffic\cite{VOD_Zipf}. 
All these factors jointly propel the concept of caching at the edge of networks\cite{video_characteristics}, 
in order to relieve the backhaul limitation and to improve end-to-end video content delivery.

Distributed caching over wireless access points has attracted a lot of attentions in the past several years. 
The authors in \cite{femoto_caching} introduced the concept of \emph{FemtoCaching} that equips femto-BSs with high storage 
capacity to store popular files. 
The content placement of distributed \emph{FemtoCaching} was shown to be NP-complete and a greedy algorithm was proposed with 
provable approximation ratio. A recent trend is to coalesce content caching with physical layer features.  
In \cite{backhaul_delay_caching}, the authors presented a caching placement algorithm to minimize the expected file down loading delay 
from a cluster of cooperative BSs equipped with cache facility. 
Cache-enabled BSs or relays can perform 
cooperative MIMO beamforming to support high-quality video streaming\cite{vincet_liu_fuck_it}. 
Authors in \cite{mobihoc:cache_MTR_ZFBF} proposed caching policies over small cell networks based on the diversity gain of maximum ratio transmission and the multiplexing gain of zero-forcing beamforming, and their objective is to maximize the average throughput of all the files. 

In this paper, we study a novel distributed caching problem of scalable videos over small cell networks. We are motivated by the 
recent prevalence of adaptive video streaming over HTTP (DASH) using scalable video coding (SVC)\cite{iDASH}.
SVC provides temporal, spatial and quality (SNR) these three dimensional scalability, 
thus each video is encoded into multiple layers consisting of a \emph{basic} layer and several \emph{enhancement} layers\cite{SVC_overview}.
The basic layer yields the minimum video quality and each level of enhancement layer provides incremental quality to 
the lower layers.
SVC allows multiple operation points (OP), 
where a sub-stream is extracted at a given bit-rates by combination of temporal, spacial and quality layers. 
Giving credit to the scalability, 
SVC is robust to channel variation, and occupies less cache space compared to the conventional dynamic adaptive video streaming\cite{SVC_test}.
When delivering SVC streaming over small cell networks, the caching placement problem becomes much more challenging.
There is a dilemma of the video QoE and cache hit ratio.
It is obviously that, storing the same video in multiple SBSs brings a higher channel diversity gain, 
while the overall hit ratio of all videos will be dragged down.
Similarly, caching a video with more layer (e.g., enhancing the video scalability), a better QoE is secured, however more cache space will be consumed.
Hence, how to allocate cache resource properly for each video over small cell networks, 
a series of interesting questions arise: 
\emph{i) which videos should be cached, 
	ii) what is the caching diversity of each video (e.g., channel diversity), 
	iii) which bit-rates of each video should be cached (e.g., video scalability), 
	so as to optimize the average QoE of streaming users, given the constraint of cache capacity ?}

To resolve these issues, we formulate an integer programming problem to maximize the average QoE of SVC streaming users 
under the constraint of cache size at each SBS. 
The average QoE is jointly determined by the distribution of video popularity, the channel characteristics and the caching 
policy. With appropriate simplifications, the original SVC caching placement problem is transformed into a multiple-choice knapsack problem (MCKP) 
solved by dynamic programming. We demonstrate the optimal scheme of caching allocation for each video, and
simulation results manifest that the proposed algorithm significantly improves the average QoE of SVC compared with the baseline algorithms.

To summarize, the major features of our work are as follows. 
\begin{itemize}
	
\item To the best of our knowledge, this is the first study on SVC caching placement over small cell networks. 
	
\item We utilize the mean opinion score (MOS) to quantify the streaming QoE, which is different from the metrics such as 
	transmission delay and average throughput in the closely related work. 
	
\item More physical layer characteristics can be further incorporated to our proactive caching model, 
as long as the connection between physical features and caching state of files is well defined.
\end{itemize}

The remainder of this paper is organized as follows. 
Section II demonstrates the system model of video transmission, caching placement and SVC QoE.
In Section III, we formulate the SVC proactive caching problem and present the proposed algorithm.
Simulation results are provided in Section IV. Finally, we conclude this paper in Section V.

\section{System Model}


\subsection{Video Transmission Model}
\begin{figure}[!t]
	\centering
	\includegraphics[width=2.7in]{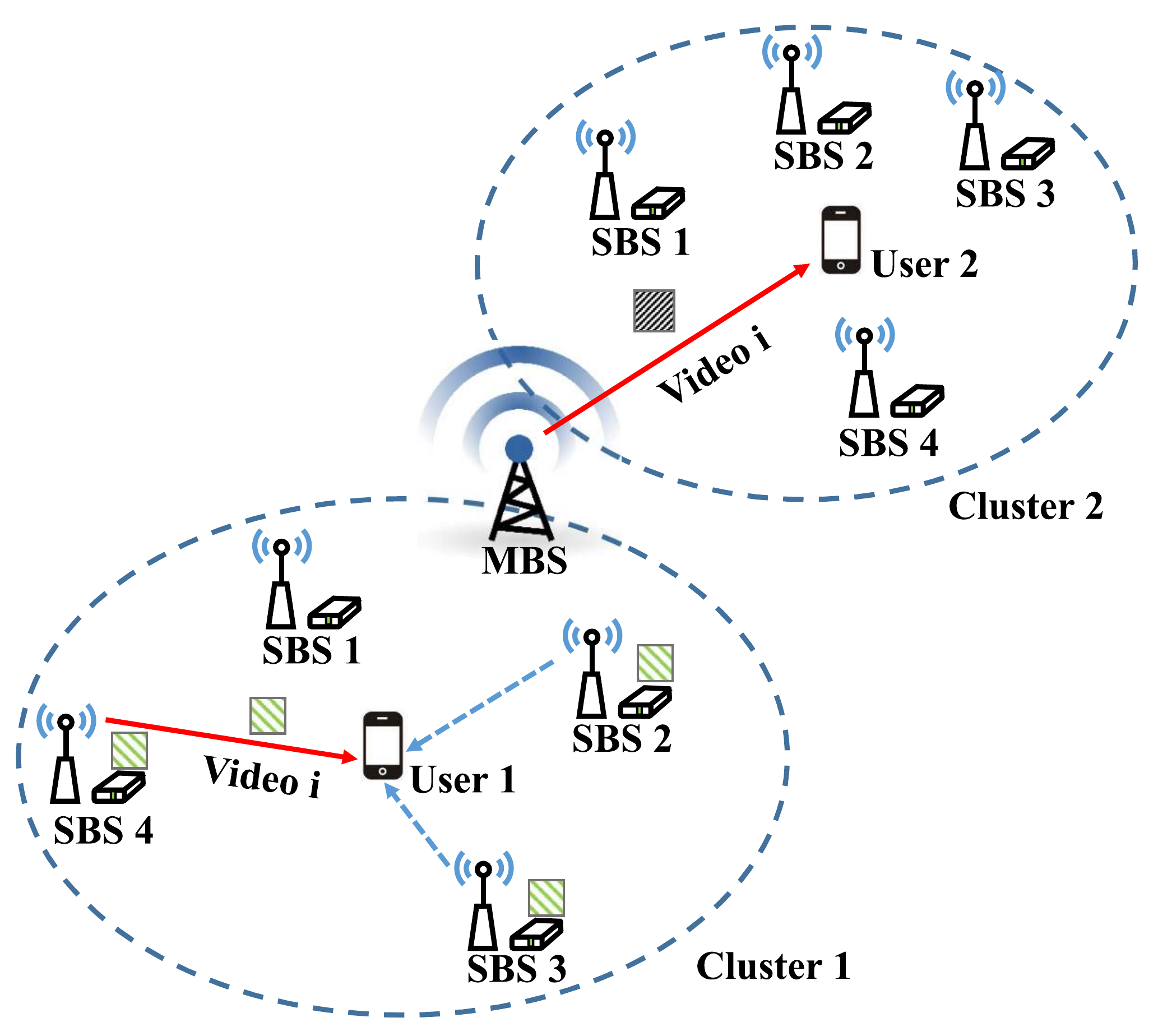}
	\caption{Video transmission and  placement model. Cluster1: video cached in SBS cluster; Cluster2: video served by MBS.}
	\label{fig1_system_model}
\end{figure}

Consider one typical video streaming user in the range of the macro cell base station (MBS). 
As shown in Fig. \ref{fig1_system_model}, each user is located at one cluster which is composed of many short range SBSs with caching ability. 
According to the prediction from content providers, some popular videos are prefetched in the local cache during the off-peak intervals.
These SBSs with cache ability will assist MBS for content delivery.

Consider the scenario where the requested video is already cached in small cells cluster.
Owing to the dense deployment, SBSs are uniformly distributed in the macro cell, 
thus we suppose that there are average $N$ SBSs in the neighborhood of a typical user. 
Let $\mathcal{N} = \{1,2,\ldots,N\}$ represent the set of SBSs.
We denote the set of candidate SBSs for transmitting video $i$ as $\Omega_i$, where $\Omega_i \subseteq \mathcal{N}$. 
As the Fig. \ref{fig1_system_model} shows, we have $\Omega_i = \{2, 3, 4\}$ in cluster 1, and $\Omega_i = \emptyset$ in cluster 2.
After acquiring the set $\Omega_i$ of potential SBSs, the user will select the SBS of the best channel quality  for video transmission.
For example, the SBS 4 is picked out from the candidate set $\{2,3,4\}$.

In another scenario, considering that the SBS is abundant in cache capacity but lacking in backhaul capcity, 
thereby, if the requested video is not cached in SBSs cluser, then the MBS will response to the user requests\cite{femoto_caching}. 
Notice that, due to the different channel state of MBS and SBSs, although the requested video ID is identical, the actual video the users experience may contain different spacial, temporal, and quality layers, which is the characteristics of SVC streaming. 
For example, in Fig. \ref{fig1_system_model}, to discern the video served by MBS or SBS, different stripe texture is marked on the sketch of video.

In each cluster, suppose users are served by a certain multiple access technology, thus we neglect the influence of interference for content placement . 
For simplicity, we assume that channel of the MBS and SBSs in one cluster seen by users, follow the same distribution.
In this paper, we consider Rayleigh fading channel, and the probability density function (PDF)  of the received SNR is given by
\begin{equation} \label{eq_rayleighFading}
P(x) = \frac{1}{\overline{SNR}}\exp \left(-\frac{x}{\overline{SNR}}\right)
\end{equation}
where $\overline{SNR}$ is the average received SNR. 
Denote the average SNR of SBSs and MBS as $\overline{\rho}$ and $\overline{\upsilon}$ respectively.
In general, we have $\overline{\rho}>\overline{\upsilon}$, because the received power from SBSs decays less within the short range cluster.

\subsection{Caching Placement Model}
Suppose that the video library contains a set of $\mathcal{M}=\{1,\cdots,M\}$ videos for proactive caching with the cardinality $M$. 
Their popularity distribution is given by the set $\mathcal{P}=\{ p_1,\cdots,p_{\scriptscriptstyle{M}} \}$ where we have $p_1\geq \cdots\geq p_M$, i.e., sorted in the descending order. 
Define $\boldsymbol{s}_i=\{0,1\}^N$, 
where $s_{i,n} = 1$ indicates that video $i$ is cached in the $n$-th SBS, and $s_{i,n} = 0$ otherwise. 
Notice that we use $l_0$-norm ${\lVert \boldsymbol{s}_i\rVert}_0 = |\Omega_i|$ to represent 
how many SBSs store the $i^{th}$ video. This norm 
can also be viewed as the degree of channel diversity for caching video $i$. 
Each SBS has a limited cache capacity denoted by $C$.

SVC steaming allows multiple operation points (OP), where a sub-stream of certain bit-rates is combined by different layers. 
Let $\mathcal{R}_i^\mathrm{\scriptscriptstyle{OP}} = \{R_i^1,\ldots,R_i^l,\ldots,R_i^{\scriptscriptstyle{L_i}}\}$ indicate the OPs of video $i$, 
where $R_i^l$ is the bit-rate of OP $l$ ($l\in \{1,2,\ldots,L_i\})$. 
A SVC streaming has better quality at a higher OP, that is, 
$R_i^1 < R_i^2 < \ldots < R_i^{\scriptscriptstyle{L_i}}$. 
We define a \mbox{2-tuple} $(\boldsymbol{s}_i, r_i)$ as the caching state of \mbox{video $i$} 
where $r_i \in \mathcal{R}_i^\mathrm{\scriptscriptstyle{OP}}$. 
The state $(\boldsymbol{s}_i, r_i)$ provides the caching details: 
\emph{video $i$ is cached in the SBSs whose $s_{i,n}=1$, and its cached bit-rates is $r_i$.} 

\subsection{QoE Model of SVC}
Content providers and operators are keen on improving user experience, which is referred to quality of experience (QoE).
According to the subjective tests of scalable video, 
it is observed that the influence of frame rate (FR) and quantization stepsizes (QS) on user perception is separable\cite{SVC_QoE_proxy_adaptation}. 
Considering a scalable video of full bit-rates $R^{\mathrm{max}}$, 
for a given sub-stream of bit-rates $r$ (at a certain OP), there exists an optimal combination of temporal, spacial and quality layers, 
which will maximize the user subjective quality.
On that account, \cite{SVC_QoE_proxy_adaptation} has derived a normalized Rate-Quality model of scalable video:  
\begin{equation}
\label{eq_normalizedQoE}
\tilde{Q}(r) = e^{-\alpha(\frac{r}{R^{\mathrm{max}}})^{-\beta}+\alpha}
\end{equation}
where $\alpha, \beta$ are model parameters, and $r\le R^{\mathrm{max}}$.

In this paper, we will consider the mean opinion score (MOS) as the metric of SVC QoE.
The MOS is expressed as a single number in range of 1 to 5, where the numbers indicate that: 
\emph{1: Bad; 2: Poor; 3: Fair; 4: Good; 5: Excellent.}\\
According to (\ref{eq_normalizedQoE}), the normalized subjective quality can be scaled as
\begin{equation}
\label{eq_mosQoE}
Q(r) = 1 + 4\tilde{Q}(r) \ .
\end{equation}

\section{QoE-Aware Caching Strategy}
In this section, our goal is to explore the optimal caching placement strategy that maximizes the QoE of SVC over small cell networks.
First, We will make connections between caching placement model and QoE of SVC, then the problem of maximizing the average QoE will be formulated.
We analyze the problem and propose a caching algorithm.

\subsection{Expected QoE of SVC Over Small Cell Networks}
According to the video transmission model we describe previously, the SBS with maximum SNR will be selected for transmission of \mbox{video $i$}, which can be represented as
\begin{equation}
\label{eq_pdfMaxSNR}
\rho_{i}^{\mathrm{\scriptscriptstyle{max}}}=\max_{k\in{\Omega_i}}\mathrm{SNR}_{i}^{k} \ .
\end{equation}
Remember that we have ${\lVert \boldsymbol{s}_i\rVert}_0 = |\Omega_i|$, thereby, according to (\ref{eq_rayleighFading}), 
the PDF of $\rho_i^{\mathrm{max}}$ is given by
\begin{equation}
\label{eq_pdfFuncMaxSNR}
P_i^{\mathrm{\scriptscriptstyle{max}}}(x) = 
\frac{{\lVert\boldsymbol{s}_i\rVert}_0}{\overline{\rho}}\exp\left(\frac{x}{\overline{\rho}}\right)\left[1-\exp\left(\frac{x}{\overline{\rho}}\right)\right]^{{\lVert\boldsymbol{s}_i\rVert}_0-1} \ .
\end{equation}

We consider the slow fading situation where the delay requirement is short compared to the channel coherence time, 
and this is also called the quasi-static scenario\cite{Tse_fundamental}. 
According to (\ref{eq_pdfFuncMaxSNR}), the outage probability of decoding OP $l$ with bit-rates $R_i^l$  can be written as
\begin{align}
\label{eq_pOutRi}
P_{\mathrm{\scriptscriptstyle{out}}}(\boldsymbol{s}_i, R_i^l)	&=\Pr\{W_{\mathrm{\scriptscriptstyle{SBS}}}\log(1+x)<R_i^l\}\notag \\
					&=\left[1-\exp\left(-\frac{2^{\frac{R_i^l}{W_\mathrm{\scriptscriptstyle{SBS}}}}-1}{\overline{\rho}}\right)\right]^{{\Vert \boldsymbol{s}_i\Vert}_0}
\end{align}
where the $W_\mathrm{\scriptscriptstyle{SBS}}$ is the average capacity allocated to a user from the selected SBS with maximum SNR.

SVC provides enhancement layers in spacial, temporal, and quality dimensions, which are successively refinable. 
Some bit rates switching mechanism exists\cite{iDASH}, which allows the server or client to adapt to the channel variation.
In general, the SVC player will first decode base layers, when the downlink reaches the threshold of an OP with higher bit-rates, 
then these enhancement layers will be decoded.
As a consequence, the expected QoE of \mbox{video $i$} under caching state $(\boldsymbol{s}_i, r_i)$ can be derived as
\begin{align}
\label{eq_expectedQoE}
\mathbb{E}\{\mathrm{QoE}{(\boldsymbol{s}_i, r_i)}\} = &\left[1-P_{\mathrm{\scriptscriptstyle{out}}}(\boldsymbol{s}_i,r_i)\right]Q(r_i)\notag \\
{}+&\! \!  \sum_{R_i^l<r_i}\!\!\!  \left[P_{\mathrm{\scriptscriptstyle{out}}}(\boldsymbol{s}_i,R_i^{l+1})-P_{\mathrm{\scriptscriptstyle{out}}}(\boldsymbol{s}_i,R_i^{l})\right]Q(R_i^l)
\end{align}
which represents that when downlink rate falls in the interval of $[R_i^l, R_i^{l+1})$, 
then the user will experience the video at the OP $l$ with bit-rates $R_i^l$. 
For simplicity, denote (\ref{eq_expectedQoE}) as $q(\boldsymbol{s}_i,r_i)$, 
which indicates the expected QoE of \mbox{video $i$} under caching state $(\boldsymbol{s}_i,r_i)$.

When the requested video is not cached in the SBS cluster, it will be delivered by  
remote servers through the MBS. The expected QoE is a special case of \eqref{eq_expectedQoE} in which 
$\mathbf{s}_i$ is substituted by $\mathbf{1}$, and $r_i$ can take any bit-rates between 
$R_i^1$ and $R_i^{\scriptscriptstyle{L_i}}$, because all the layers are stored in the remote servers. 
Similarly, the SNR $\bar{\rho}$ and bandwidth $W_{\mathrm{\scriptscriptstyle{SBS}}}$ of the SBS are replaced 
by corresponding variables of the MBS, i.e. $\bar{v}$ and $W_{\mathrm{\scriptscriptstyle{MBS}}}$. 
With certain abuse of notation, we denote $q^{\mathrm{\scriptscriptstyle{MBS}}}(R_i^{\scriptscriptstyle{L_i}})$ as the average QoE of 
video $i$ served by MBS. Therefore, the expected QoE of all the videos over small cell and macro cell networks is given by
\begin{eqnarray}
\label{eq_optFunc}
Q_{\mathrm{\scriptscriptstyle{SCMC}}}=\sum_{i=1}^{M}p_i\left[\mathbf{1}_{(\boldsymbol{s}_i\neq 0)}q(\boldsymbol{s}_i,r_i)
+\mathbf{1}_{(\boldsymbol{s}_i = 0)}q^\mathrm{\scriptscriptstyle{MBS}}(R_i^{\scriptscriptstyle{L_i}})\right]
\end{eqnarray}
where $\boldsymbol{1}_{(\boldsymbol{s}_i\neq 0)}$ and $\boldsymbol{1}_{(\boldsymbol{s}_i = 0)}$ are indication functions.

\subsection{Problem Formulation and Analysis}
Our objective is to maximize the overall QoE by proactively caching videos at each SBS.
Denote by \mbox{$\mathcal{T}=\{t_1,t_2,\ldots,t_M\}$} the set of video durations. 
The optimal content placement problem for scalable videos can be formulated as:
\begin{align}
\label{eq_optimization1}
\mathcal{P}_0: \underset{(\boldsymbol{s}_i,r_i)}{\mathrm{maximize}} \quad  & Q_{\mathrm{\scriptscriptstyle{SCMC}}}   \\
\textrm{subject to} \quad & \sum_{i=1}^{M}r_i t_i s_{i,n} \leq C, &n =1,\ldots,N \tag{\ref{eq_optimization1}.a}\\
\quad &\boldsymbol{s}_i\in\{0,1\}^N,&i=1,\ldots,M \tag{\ref{eq_optimization1}.b}\\
\quad &r_i \in \mathcal{R}_i^{\mathrm{\scriptscriptstyle{OP}}}, &i=1,\ldots, M \tag{\ref{eq_optimization1}.c}
\end{align}
where (\ref{eq_optimization1}.a) is the cache size constraint of each SBS.

The optimization problem $\mathcal{P}_0$ is an integer programming, which is in general very 
complicated to be solved. The computational complexity grows exponentially with regard to 
the diverse video durations, bit-rate levels and number of SBSs. To determine the 
optimal caching state $(\boldsymbol{s}_i,r_i)$ for each video, it is necessary to 
make certain reasonable simplifications.

Firstly, we assume that all the videos have the same number of OPs, and the bit-rates of each OP is identical, i.e. 
$\mathcal{R}_1^{\mathrm{\scriptscriptstyle{OP}}}=\mathcal{R}_2^{\mathrm{\scriptscriptstyle{OP}}}=\ldots=\mathcal{R}_{\scriptscriptstyle{M}}^{\mathrm{\scriptscriptstyle{OP}}}=\{R^1, R^2,\ldots,R^{\scriptscriptstyle{L}}\}$. 
This is a natural assumption because the content providers usually follow certain 
standards to encode the raw videos. 
We further assume that all the videos have the same duration denoted by $T$. This 
assumption does not change the original problem in \eqref{eq_optimization1}. 
For a video longer than $T$, it can be regarded as several 
videos with the same duration $T$ and popularity. 

Eq. \eqref{eq_pOutRi} and $\mathcal{P}_0$ indicate that the number of candidate SBSs 
that have cached the requested video will influence the expected QoE. 
Denote \mbox{$n_i={\rVert\boldsymbol{s}_i\lVert}_0$}, where $n_i$ stands for the number of SBSs that store the $i^{th}$ video. 
Hence, we only need to determine the caching state $(n_i,r_i)$ 
instead of $(\boldsymbol{s}_i, r_i)$. 
Then, $P_{\mathrm{\scriptscriptstyle{out}}}(\boldsymbol{s}_i, r_i)$ in \eqref{eq_expectedQoE} can be rewritten as $P_{\mathrm{\scriptscriptstyle{out}}}(n_i, r_i)$,
and for simplicity, we denote $q(n_i,r_i)$ as the average QoE of video $i$ under caching state $(n_i,r_i)$.
With above operations, the vector of constraints on the cache size can be merged together. 
We can deem it as that all the $N$ SBSs in the cluster are treated as a big cache 
with the size of $NC$. Hence, the constraint (\ref{eq_optimization1}.a) is relaxed to
\begin{equation*}
\sum n_i r_i \leq N\widetilde{C}
\end{equation*}
where $\widetilde{C}$ is the normalized cache size, calculated by the lowest bit rates $R^1$ and duration $T$.
Since we take all the SBSs as an entirety and skip the process of allocating videos to specific SBS, 
there may exit a few videos  which can not be cached in the residual space of a single cache. 
However, even one SBS cache with 1 TB can store hundreds videos, 
thus the number of these outliers can be ignored compared to those videos which are already cached.
Moreover, owing to the scalability of SVC, we can set these exceptional videos at a lower bit rates.
To sum up, the change in the constraint is trivial to results of proactive caching.

According to the above analysis, the problem $\mathcal{P}_0$ evolves into determining caching state $(n_i, r_i)$ of each video, where $n_i\in\{1,2,\ldots,N\}$, and $r_i\in\{R^1,R^2,\ldots,R^{\scriptscriptstyle{L}}\}$.
The total number of combinations of \mbox{$n_i$ and $r_i$} is $NL$, but some combinations are obviously inefficient. 
Since all videos are encoded with the same number of OPs and bit-rates, we omit the subscript $i$ 
when considering the combinations of $n_i$ and $r_i$. 
Let $q^{\mathrm{\scriptscriptstyle{MBS}}}$ denote $q^{\mathrm{\scriptscriptstyle{MBS}}}(R_i^{\scriptscriptstyle{L}})$,
and it is a constant for all videos given these assumption above. 
Consequently, the efficient caching state must obey the following criteria
\begin{cachingState}
\label{props_caching_state}
Cache space should be allocated based on the efficient combinations of $n_i$ and $r_i$, which satisfy that
\begin{align}
\label{efficient_caching_state}
&q^{\mathrm{\scriptscriptstyle{MBS}}}<q(n^{\scriptscriptstyle{(1)}}, r^{\scriptscriptstyle{(1)}})\leq q(n^{\scriptscriptstyle{(2)}},r^{\scriptscriptstyle{(2)}})\leq \ldots \leq q(n^{\scriptscriptstyle{(V)}},r^{\scriptscriptstyle{(V)}})  \notag  \\
&n^{\scriptscriptstyle{(1)}}r^{\scriptscriptstyle{(1)}}\leq n^{\scriptscriptstyle{(2)}}r^{\scriptscriptstyle{(2)}}\leq \ldots \leq n^{\scriptscriptstyle{(V)}}r^{\scriptscriptstyle{(V)}}
\end{align}
where the superscript indicates the ascending order of expected QoE.
Denote $\mathcal{CS}=\{(n^{\scriptscriptstyle{(1)}},r^{\scriptscriptstyle{(1)}}),\ldots,(n^{\scriptscriptstyle{(V)}},r^{\scriptscriptstyle{(V)}})\}$ as set of efficient caching state, 
and $(n_i,r_i)\in\mathcal{CS}$, for any cached video.
Moreover, let $\mathcal{Q}=\{q(n^{\scriptscriptstyle{(1)}},r^{\scriptscriptstyle{(1)}}),\ldots,q(n^{\scriptscriptstyle{(V)}},r^{\scriptscriptstyle{(V)}}) \}$ and
$\mathcal{C}=\{n^{\scriptscriptstyle{(1)}}r^{\scriptscriptstyle{(1)}},\ldots,n^{\scriptscriptstyle{(V)}}r^{\scriptscriptstyle{(V)}}\}$,
which denote the efficient set of QoE reward and set of cache consumption respectively.
\end{cachingState}

\begin{IEEEproof}
\label{proof_cahing_state}
	Assume an arbitrary caching state $(n',r')$, whose  $q(n',r') \leq q^\mathrm{\scriptscriptstyle{MBS}}$.
Apparently, the cache consumption $n'r'$ is not a necessity, 
because the video under caching state $(n',r')$ can be delivered by MBS with better QoE, 
and the saved cache space will be utilized to enhance the QoE of other videos.

Similarly, consider a caching state $(n'',r'')$, whose QoE satisfies $q(n^{\scriptscriptstyle{(k)}},r^{\scriptscriptstyle{(k)}})\leq q(n'',r'') \leq q(n^{\scriptscriptstyle{(k+1)}},r^{\scriptscriptstyle{(k+1)}})$.
If the cache consumption satisfies $n^{\scriptscriptstyle{(k)}}r^{\scriptscriptstyle{(k)}}\leq n''r'' \leq n^{\scriptscriptstyle{(k+1)}}r^{\scriptscriptstyle{(k+1)}}$, then $(n'',r'')$ is a efficient  state.
When $n''r''> n^{\scriptscriptstyle{(k+1)}}r^{\scriptscriptstyle{(k+1)}}$, then combination $(n^{\scriptscriptstyle{(k+1)}},r^{\scriptscriptstyle{(k+1)}})$ is superior to $(n^{''},r^{''})$ with less cache consumption and higher QoE, consequently $(n'',r'')$ is not an efficient state.
While $n''r''<n^{\scriptscriptstyle{(k)}}r^{\scriptscriptstyle{(k)}}$, state $(n'',r'')$ is more efficient than $(n^{\scriptscriptstyle{(k)}},r^{\scriptscriptstyle{(k)}})$,
therefore, substitute state $(n'',r'')$ for $(n^{\scriptscriptstyle{(k)}},r^{\scriptscriptstyle{(k)}})$.

Follow  the procedure above, find out all the qualified efficient state, and set $\mathcal{CS}$ will be determined.
The set $\mathcal{CS}$ of efficient caching state, 
makes a mapping from set $\mathcal{C}$ of cache consumption  to set $\mathcal{Q}$ of QoE reward.
\end{IEEEproof}
After acquiring the efficient caching state, problem $\mathcal{P}_0$ is reduced to selecting a caching state from $\mathcal{CS}$ 
for videos with different popularity.
An important question is how the cache space is 
allocated to each video.
Proposition \ref{props_caching_state} and \eqref{eq_optFunc} jointly yield the following proposition: 
\begin{spaceStairs}
\label{props_space_stairs}
For the optimality of problem $\mathcal{P}_0$, video with higher popularity will be endowed with more cache space, 
compared to those with lower popularity, which can be described as
\begin{align}
\label{eq_cacheStairs}
&q(n_1,r_1)\geq \ldots q(n_i,r_i)\geq \ldots \geq q(n_{\scriptscriptstyle{M}}, r_{\scriptscriptstyle{M}}) \notag \\
&n_1r_1\geq \ldots \geq  n_ir_i \geq \ldots \geq n_{\scriptscriptstyle{M}}r_{\scriptscriptstyle{M}}
\end{align}
where $(n_i,r_i)$ is the optimal caching state of video $i$.
\end{spaceStairs}

\begin{IEEEproof}
\label{proof_space_stairs}
Consider \mbox{video $a$} and \mbox{video, $b$}, whose popularity satisfy that $p_a>p_b$, 
and their related cache state are $(n_a,r_a)$ and $(n_b,r_b)$.
The sum QoE of these two videos is ${p_a}q(n_a,r_a)+{p_b}q(n_b,r_b)$.
If $n_ar_a<n_br_b$, then swap the cache space of video $a$ and video $b$, 
then the new QoE is ${p_a}q(n_b,r_b)+{p_b}q(n_a,r_a)$.
The difference value between the new and old QoE is
$$(p_a-p_b)(q(n_b,r_b)-q(n_a,r_a))>0$$
thereby, higher QoE will be obtained by allocating more cache space to videos with higher popularity.
Intuitively, in order to maximize the overall QoE ,
the optimal structure of cache space allocation for each video can be shaped as descending stairs according to the descending popularity. 
\end{IEEEproof}

Based on Proposition \ref{props_caching_state} and Proposition \ref{props_space_stairs}, 
problem $\mathcal{P}_0$ can be transformed to
\begin{align}
\label{eq_optimization2}
\mathcal{P}_1:\ \underset{\hat{m},\ (n_i,r_i)}{\mathrm{maximize}} \quad &\sum_{i=1}^{\hat{m}}p_iq(n_i,r_i)+\sum_{i=\hat{m}+1}^{M}p_iq^{\mathrm{\scriptscriptstyle{MBS}}} \\
\textrm{subject to} \quad &\sum_{i=1}^{\hat{m}}n_ir_i \leq N\widetilde{C} \tag{\ref{eq_optimization2}.a} \\
& \hat{m} \in \{1,\ldots,M\} \tag{\ref{eq_optimization2}.b} \\
& (n_i,r_i)\in \mathcal{CS},\ i=1,\ldots,\hat{m} \tag{\ref{eq_optimization2}.c} 
\end{align}
where $\mathcal{CS}=\{(n^{\scriptscriptstyle{(1)}},r^{\scriptscriptstyle{(1)}}),\ldots,(n^{\scriptscriptstyle{(V)}},r^{\scriptscriptstyle{(V)}})\}$. 
The solution to problem $\mathcal{P}_1$ determines: i) 
$\hat{m}$: videos from $1 \textrm{ to } \hat{m}$ will be cached in SBS cluster;
ii) $(n_i,r_i)$: the optimal state of each cached video.

\subsection{Proposed Algorithm}
Problem $\mathcal{P}_1$ can be viewed as a multiple-choice knapsack problem (MCKP)\cite{MCKP_algorithm}, 
in which $M$ videos with various popularities are viewed as different items, 
and the SBS cluster is treated as a knapsack with the volume $N\widetilde{C}$.
The QoE reward set $\mathcal{Q}$ and cache consumption set $\mathcal{C}$ are obtained by determining efficient caching state $\mathcal{CS}$,
as a consequence, each video has total $V$ choices of cache state. 
According to (\ref{eq_optimization2}), let $F(m,v)$ denotes the overall QoE under 
the state $(m,v)$, 
where $m$ is the number of candidate videos, $v$ is the total cache size of the SBS cluster.
The iteration relation of $F(m,v)$ and $F(m-1,v)$ can be described as
\begin{align}
\label{eq_iteration}
F(m,v)=&\max\{F(m-1,v)+p_m{q^{\mathrm{\scriptscriptstyle{MBS}}}}, \notag \\
&F(m-1,v-n_mr_m)+p_m{q(n_m,r_m)} \}, \notag \\
&(n_m,r_m)\in\mathcal{CS}, \textrm{ \emph{for all cached video.}}
\end{align}
Whether the $m^{th}$ video is cached in SBS cluster or not depends on the QoE reward brought by adding the $m^{th}$ video. 
If the reward of caching the $m^{th}$ video does not surpass the reward brought by MBS, 
video $m$ will be served by the MBS instead of being cached in the SBS cluster.
The MCKP can be solved by dynamic programming using the iteration relation \eqref{eq_iteration} in pseudo-polynomial time. 
Thereby, we present \mbox{\textbf{Algorithm \ref{alg_MCKP}}} to determine $\hat{m}$ and 
$(n_i,r_i)$ of each video.
The running time of \mbox{\textbf{Algorithm \ref{alg_MCKP}}} is $O(\hat{m}N\widetilde{C}V)$.
\begin{algorithm}
	\caption{Determine $\hat{m}$ and $(n_i,r_i)$ by solving MCKP}
	\label{alg_MCKP}
	\begin{algorithmic}
		\STATE 
		According to Proposition \ref{props_caching_state}, determine efficient caching state $\mathcal{CS}$,
		and relevant QoE set $\mathcal{Q}$, and consumption set $\mathcal{C}$ 
		\FOR{$m=1$ \TO $M$}
		\FOR{$v=1$ \TO $N\widetilde{C}$}
		\STATE
		$F(m,v)=\max\{F(m-1,v)+p_m{q^{\mathrm{\scriptscriptstyle{MBS}}}},$\\
		$\quad \quad \quad \ \ F(m-1,v-n_mr_m)+p_m{q(n_m,r_m)} \}$, \\
		where state $(n_m,r_m)\in\mathcal{CS}$
		\STATE Update caching  state of each video 
		\ENDFOR
		
		\IF{video $m$ is served by MBS, given cache size of $N\widetilde{C}$}
		\STATE
		$\hat{m}=m-1$
		\STATE
		Videos beginning with $m$ are served by MBS
		\RETURN $\hat{m}$ and $(n_i,r_i)$, where $i=1,\ldots,\hat{m}$
		\ENDIF
		\ENDFOR
		\STATE
		$\hat{m}=M$\\
		\RETURN
		$\hat{m}$ and $(n_i,r_i)$, where $i=1,\ldots,\hat{m}$
	\end{algorithmic}
\end{algorithm}

\section{Simulation Results}
In this section, we present numerical results of the QoE-aware proactive caching of scalable videos over small cell networks.

The content provider offers a library of $M=10000$ candidate videos.
Assume that the popularity of these videos follow Zipf distribution with shape parameter $s=0.8$\cite{VOD_Zipf}.
The parameters for SVC QoE are set as $\alpha = 0.16, \ \beta = 0.66$ according to \cite{SVC_QoE_proxy_adaptation}.
We set the average duration is $T=1$ hour, 
and the specifics of SVC we experiment on is elaborated in Table \ref{table_video}, 
according to the suggestions from \cite{SVC_test}.
We demonstrate three sets of experimental results as below.
First, the structure of optimal caching strategy is highlighted, revealing the optimal caching state of each video. 
Next, we illustrate the tendency of average QoE and hit ratio, under the different number of SBSs and size of cache space.
Finally, we compare the performance of our proposed caching strategy with other reference strategies.

\subsection{Caching State of Scalable Video}
\begin{figure*}[!t]
	\centering
	\subfloat[Degree of caching diversity (channel diversity) for each video]{\includegraphics[width=2.1in]{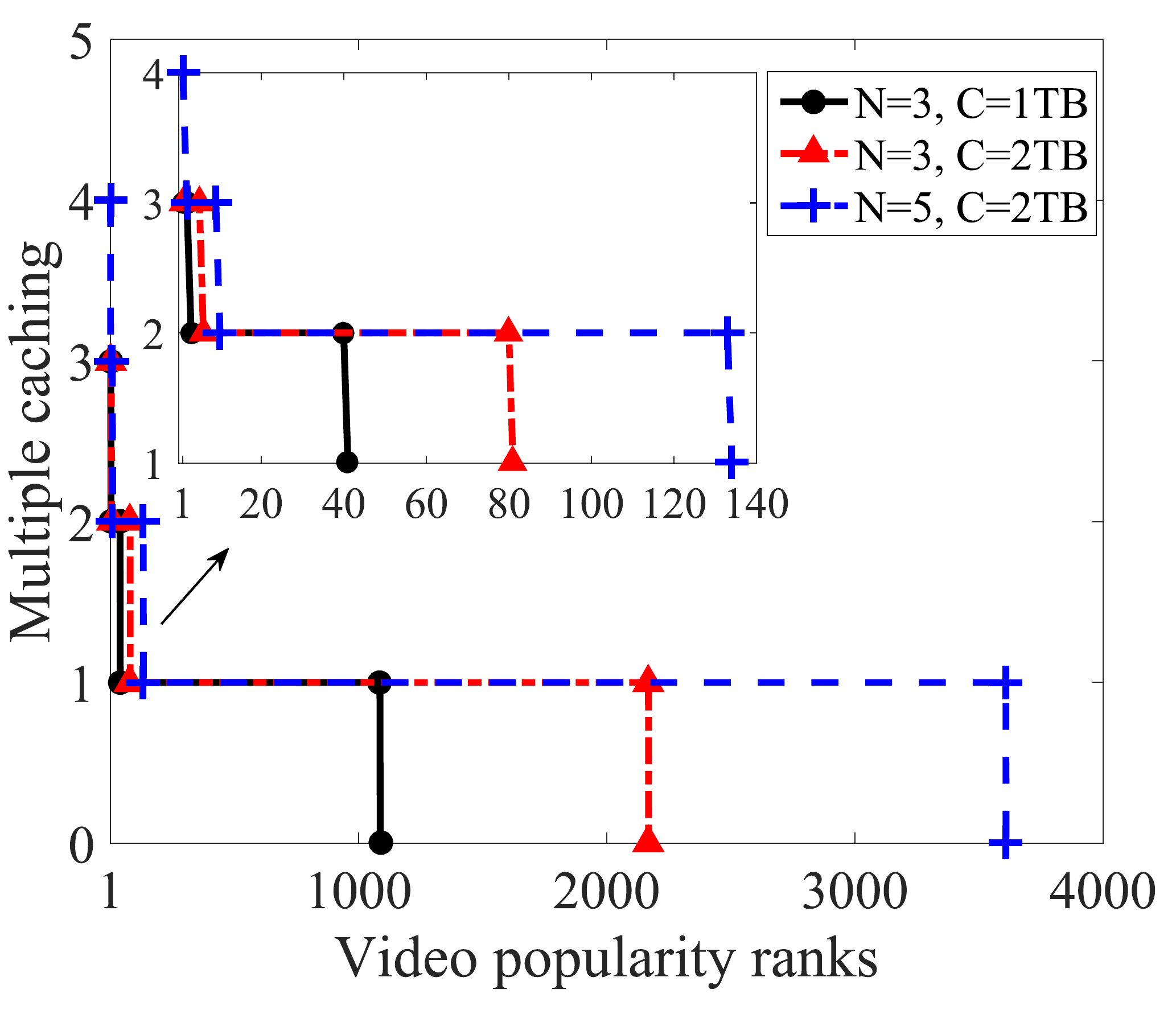}
	\label{fig_multiple_caching}}
	\hfil
	\subfloat[Bit rates selected for each video (video scalability)]{\includegraphics[width=2.1in]{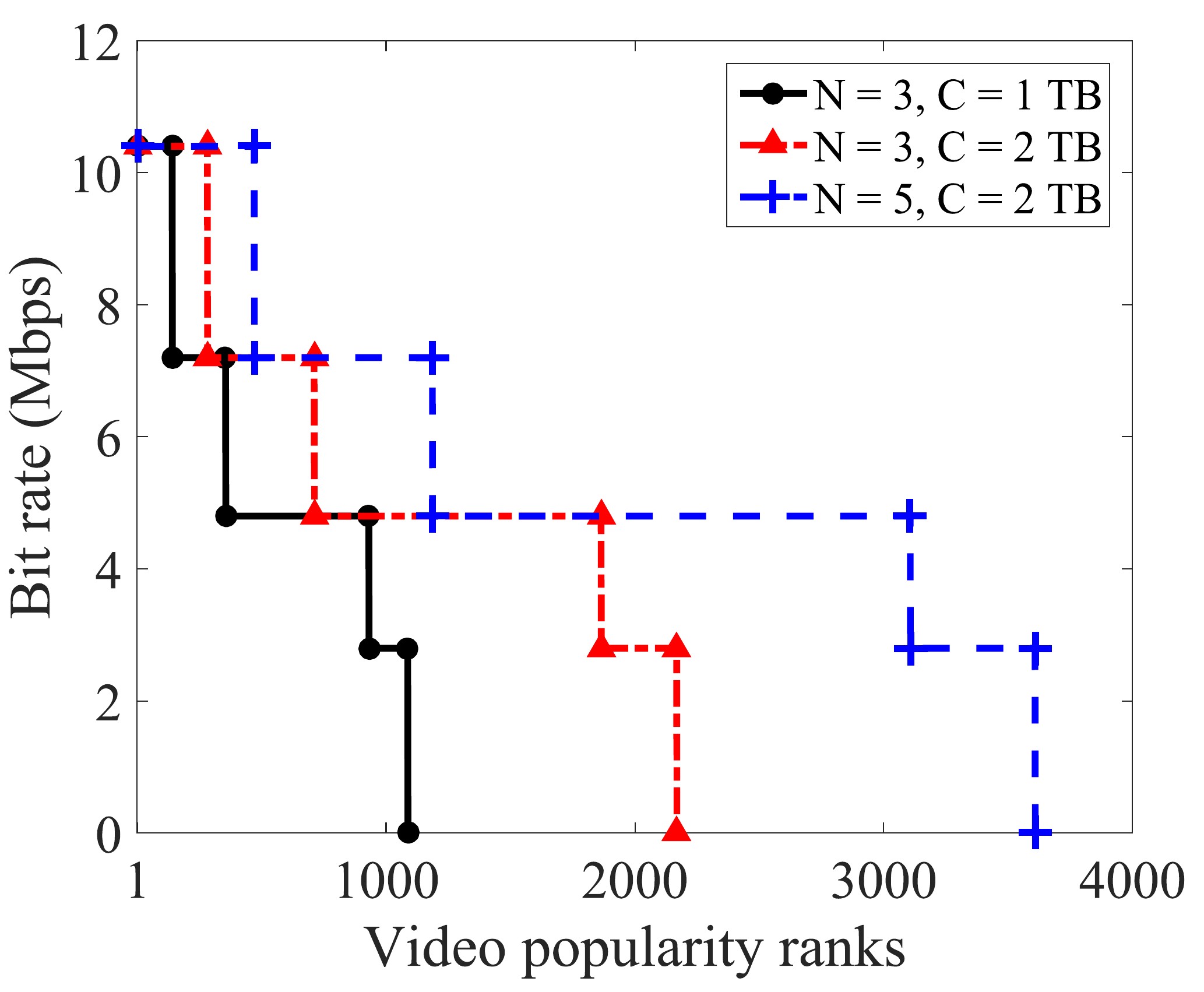}
	\label{fig_bitrate_state}}
	\hfil
	\subfloat[QoE of each cached videos]{\includegraphics[width=2.1in]{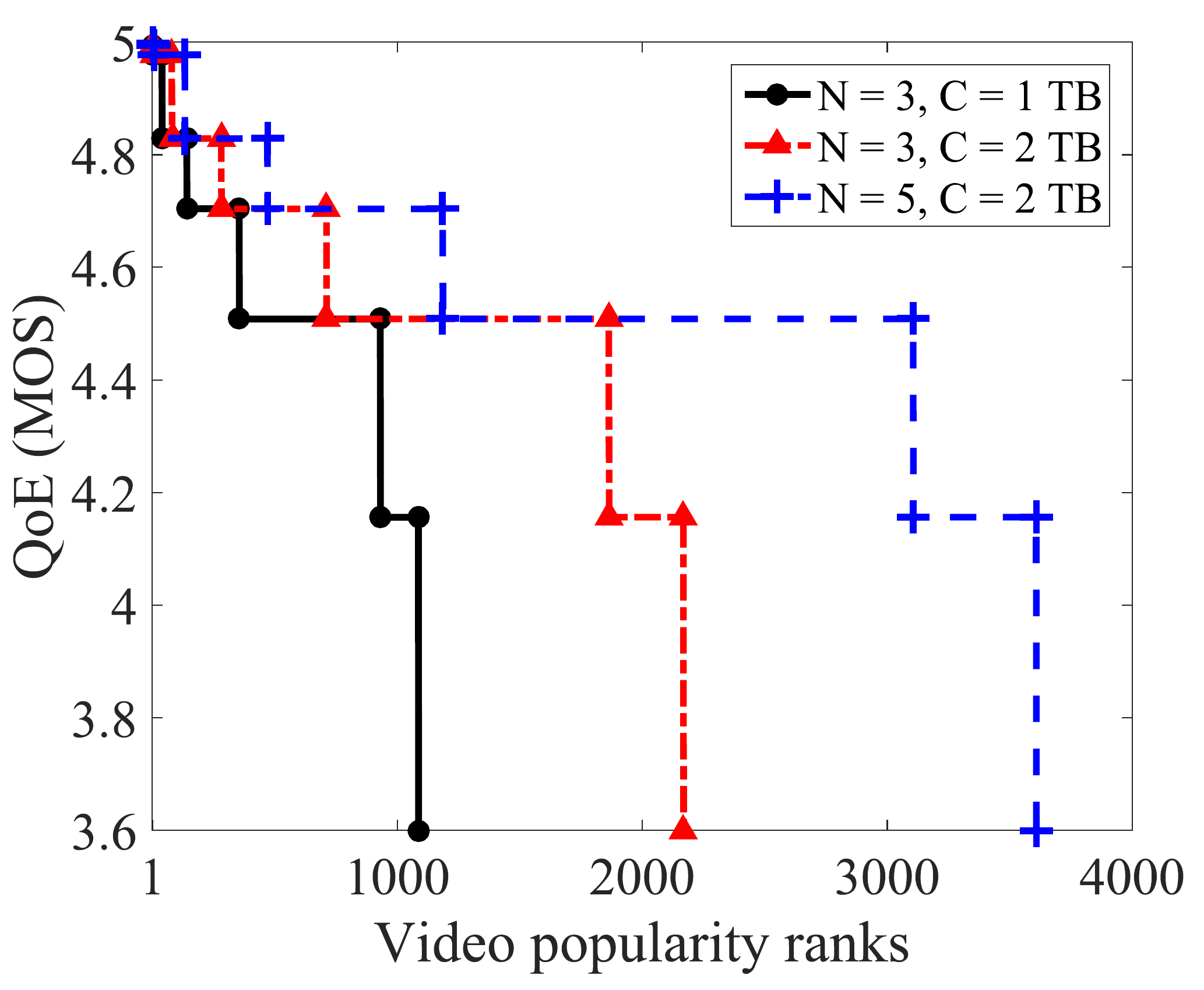}
	\label{fig_QoE_state}}
	\caption{Caching state of scalable videos with different popularity. $\overline{\rho}=10\textrm{ dB}$, $W_{\mathrm{\scriptscriptstyle{SBS}}}=5\textrm{ MHz}$; $\overline{\upsilon}=3{\textrm{ dB}}$, $W_{\mathrm{\scriptscriptstyle{MBS}}}=2{\textrm{ MHz}}$}
	\label{fig_caching_state}
\end{figure*}

The averaged SNR and shared bandwidth for a user from the MBS are set to $\overline{\upsilon}=3{\textrm{ dB}}$ and $W_{\mathrm{\scriptscriptstyle{MBS}}}=2{\textrm{ MHz}}$.
Similarly, the averaged SNR and bandwidth provided by SBS are set as $\overline{\rho}=10\textrm{ dB}$ and $W_{\mathrm{\scriptscriptstyle{SBS}}}=5\textrm{ MHz}$. 
We consider several scenarios where the number of SBSs is $N = 3$ or $N = 5$ in one cluster, and the cache size of each SBS is $C = 1 \textrm{ or } 2 \textrm{ TB}$ respectively.
We next show the caching state of each scalable video with different popularities. 
Fig.(\ref{fig_multiple_caching}) represents the degree of caching diversity of hot videos. 
The top one hundred videos are repeatedly cached in more than one SBS, 
which can provide diversity gain for these hot videos to achieve a higher throughput. 
Meanwhile, we notice that the unpopular videos are not cached. 
Fig.(\ref{fig_bitrate_state}) demonstrates the selected bit rates of these cached videos. 
The popular ones are allocated with higher bit-rates, thus possessing more OPs to 
adapt to channel variation and achieving better QoE.
It is worth observing that the caching diversity falls much more faster than video scalability, as the popularity decreases.
Given the caching state $(n_i, r_i)$ of each video represented in Fig.(\ref{fig_multiple_caching}) and Fig.(\ref{fig_bitrate_state}), we plot the QoE of each cached video in Fig.(\ref{fig_QoE_state}). One can observe that the MOS QoE of a video decreases along with its popularity. 
\begin{table}[!t]
	\renewcommand{\arraystretch}{1.3}
	\caption{Specifics of SVC Streaming}
	\label{table_video}
	\centering
	\begin{tabular}{|c||c||c|}
		\hline
		\bfseries Resolution & \bfseries Suggested bit rates (Mbps) &\bfseries Operation points \\
		\hhline{|=#=#=|}
		$1920\times1080$ & $10.4,\ 7.2$ & $R^{10}$, $R^9$ \\
		\hline
		$1280\times720$  & $\ \, 4.8,\  2.8$  & $R^8, \ R^7$ \\
		\hline
		$960\times540$ & $\ \,2.0, \  1.2$ & $R^6,  \ R^5$ \\
		\hline
		$640\times360$ & $\ \,1.0, \  0.6$ & $R^4, \ R^3$ \\
		\hline
		$352\times288$ & $\  \,0.3$ & $R^2$ \\
		\hline
		$176\times144$ & $\ \,0.1$ & $ R^1$ \\
		\hline
	\end{tabular}
\end{table}

\subsection{Tendency of QoE and Hit Ratio}
\begin{figure}[!t]
\centering
\subfloat[Average QoE of SBS cluster]{\includegraphics[width=1.6in]{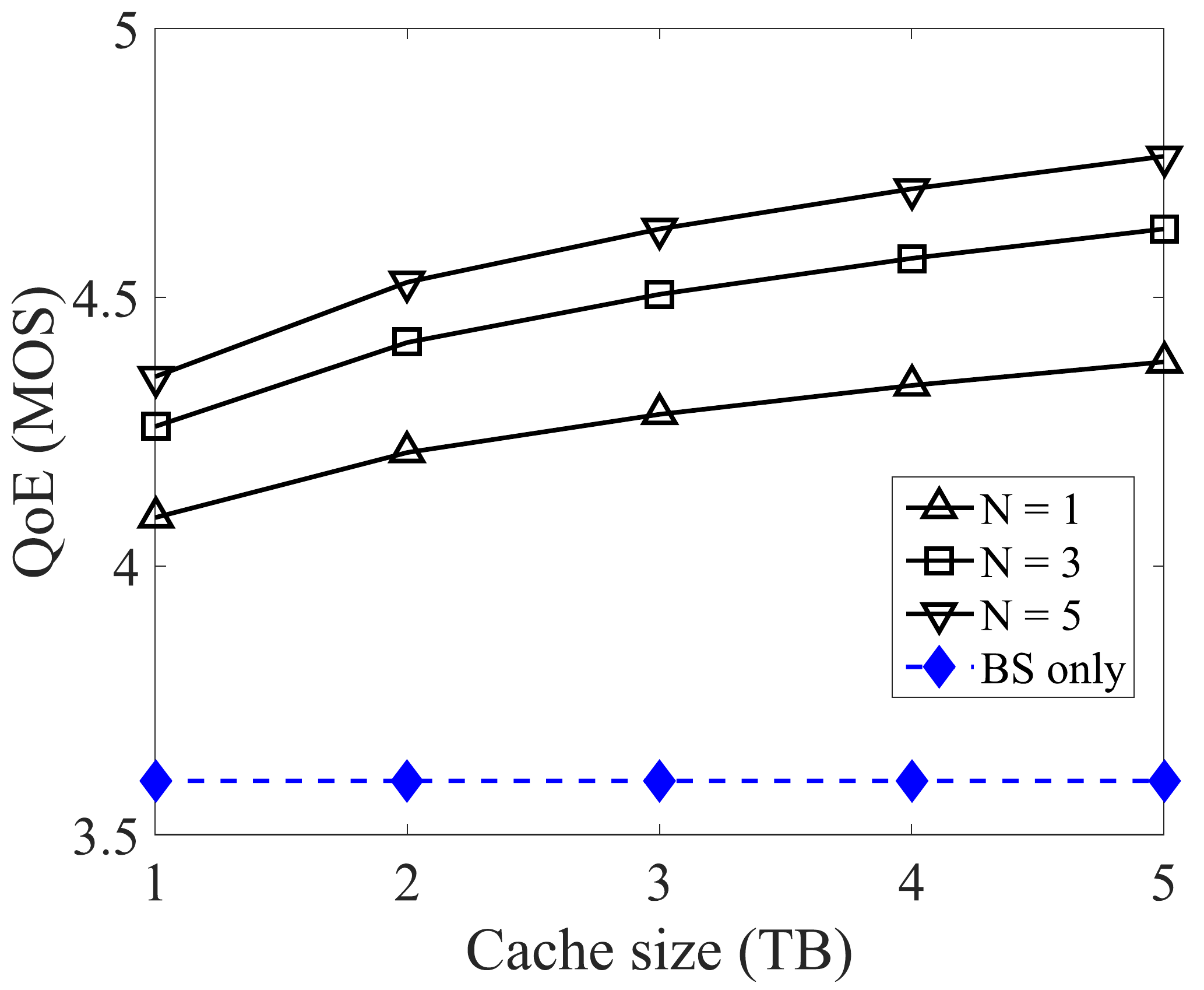}
\label{fig_qoe}}
\subfloat[Overall hit ratio of SBS cluster]{\includegraphics[width=1.6in]{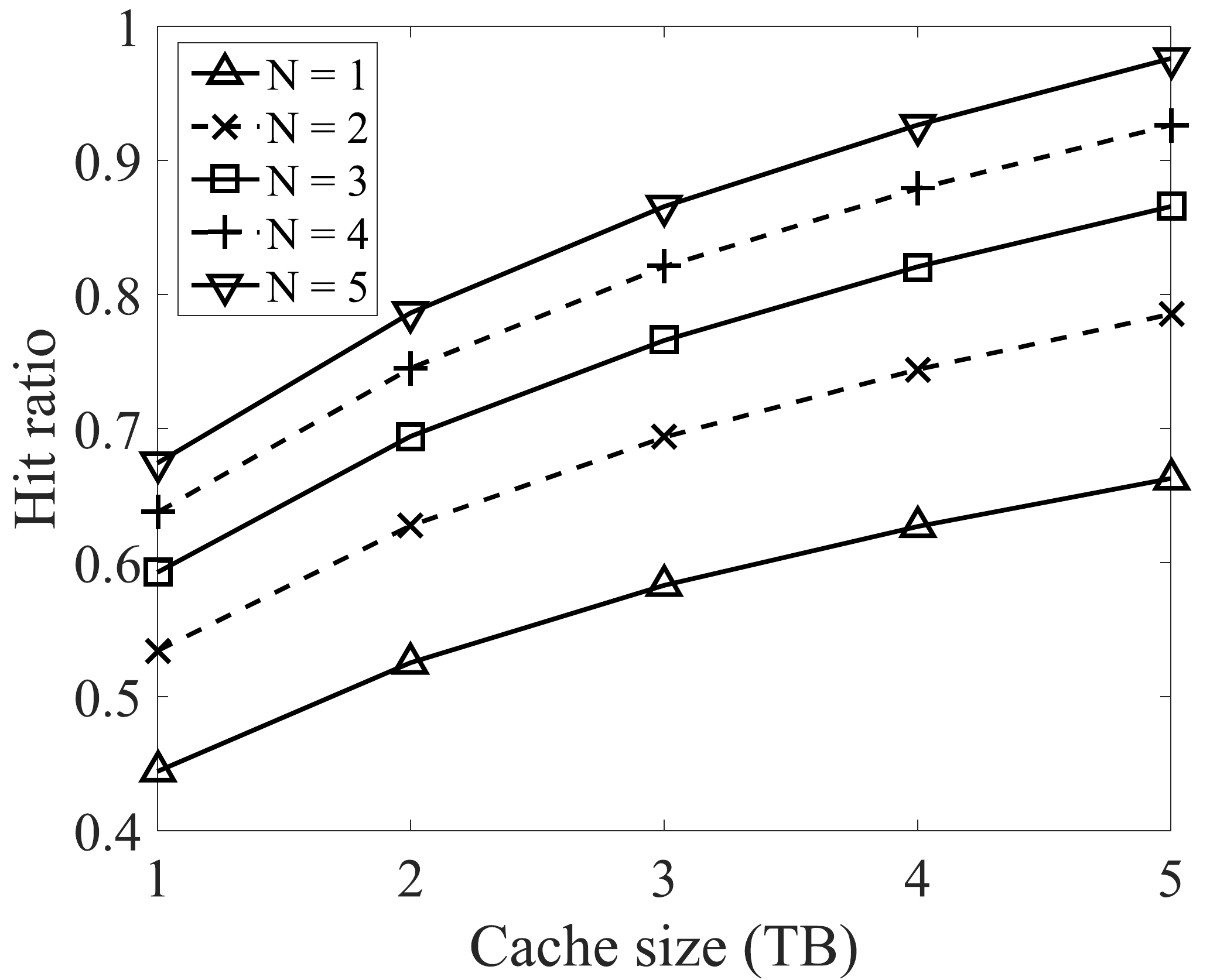}
\label{fig_hitratio}}
\caption{Tendency of QoE and hit ratio under various scenarios}
\label{fig_qoe_hitratio}
\end{figure}

We now proceed to explore the tendency of QoE and hit ratio
under different cache size and different number of SBSs. 
The channel condition is the same as that of the previous experiments.
Fig.(\ref{fig_qoe}) shows that even one SBS cache with 1 TB will boost the average 
QoE compared to the scenario where there is only MBS. 
Higher average QoE can be provided by adding number of SBSs in cluster than simply increasing the 
space of a single cache, owing to the additional diversity gain brought by multiple SBSs.
In spite of the top 100 or 200 videos cached in more than one SBSs, the majority of videos are cached only in one SBS, due to the heavy tail property of Zipf distribution, thus the overall hit ratio mainly depends on the total cache size of the SBS cluster, 
which is illustrated in Fig.(\ref{fig_hitratio}).

\subsection{Comparison with Other Caching Strategies}
\begin{figure}[!t]
	\centering
	\includegraphics[width=2.6in]{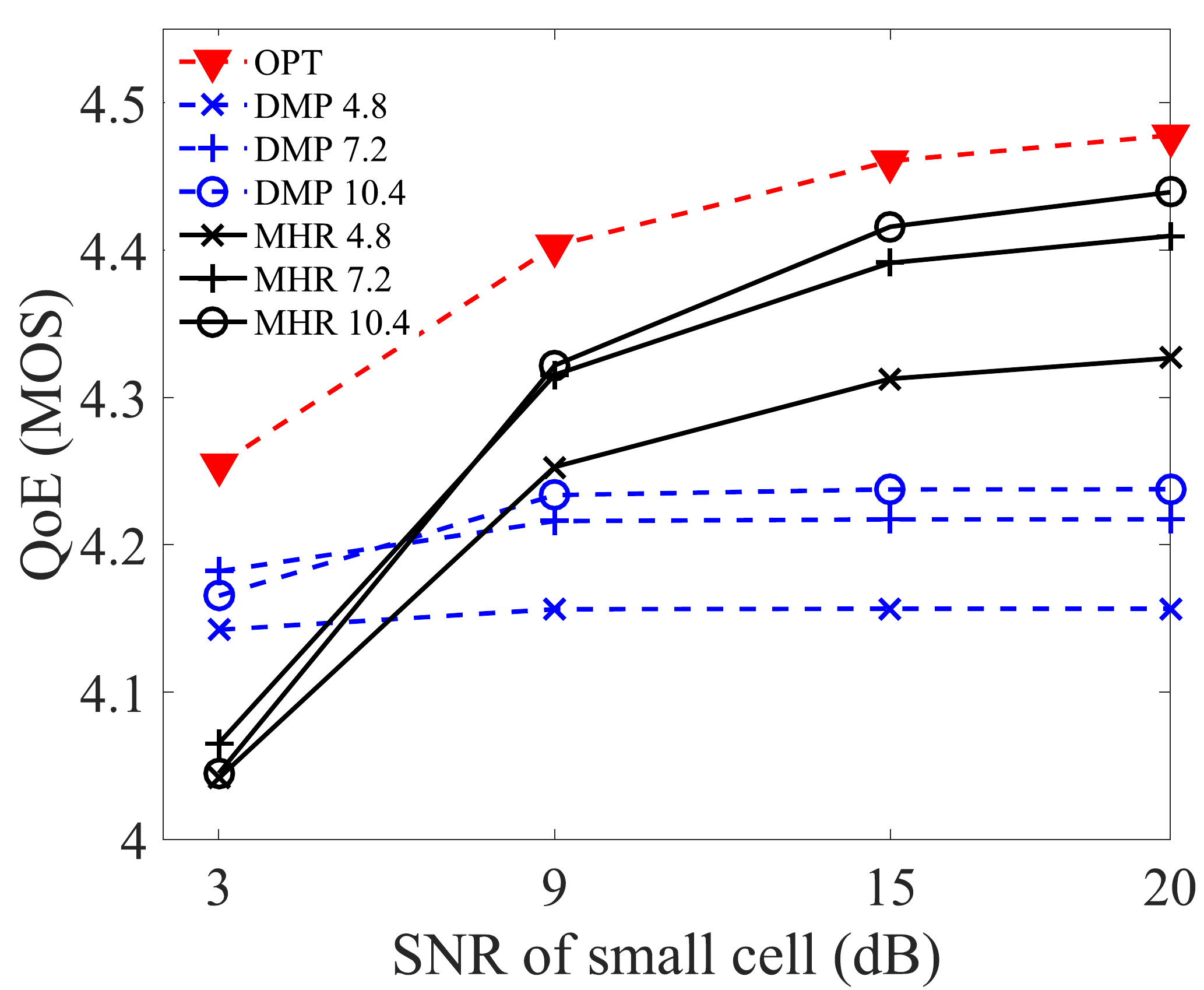}
	\caption{QoE comparison between baseline caching strategies and the proposed strategy ($N=3,C=2\ \mathrm{TB}$).}
	\label{fig_comparison}
\end{figure}

Next, we compare the proposed caching strategy with two baseline caching schemes.
We consider a representative circumstance when there are three SBSs in the neighborhood of a user, and the cache size of each SBS is 2 TB.
The first reference caching strategy duplicates the most popular videos in each SBS cache (DMP), 
thus the cache content of any SBS is identical. 
The other baseline scheme only caches one copy of each video in the SBS cluster in order to maximize the hit ratio (MHR).
We consider three types of bit-rates selected for baseline strategies, i.e., 4.8 Mbps, 7.2 Mbps, and 10.4 Mbps respectively.
Fig.(\ref{fig_comparison}) demonstrates the average QoE of these caching strategies 
under various SNRs of SBS. We observe that DMP outperforms MHR in the low SNR scenarios, but has 
poor performance compared with MHR in the high SNR scenarios.
The proposed caching strategy is superior to baseline strategies in all the situations. 
The reason lies in that it considers the diversity gain of multiple caching for hot videos, 
and adjusts the cached bit-rates adaptively for the videos with different popularities.

\section{Conclusion}
This paper investigated the proactive caching strategy for maximizing the average QoE 
of SVC over small cell networks. 
Caching videos in collocated small cell base stations may not only reduce transmission range, but 
also bring channel diversity gains. 
We formulate an integer programming problem for the optimal SVC placement, given the 
distribution of video popularity, storage constraints and wireless channel characteristics. 
The SVC caching placement is relaxed as a multiple-choice knapsack problem (MCKP) 
to reduce the computational complexity. 
The proposed caching algorithm determines the set of videos to be cached, plus the configuration details of them, 
i.e., the selected bit-rates (video scalability) and caching diversity (channel diversity) for each video. 
Simulation results demonstrate that our algorithm significantly outperforms two baseline 
caching strategies in terms of the average QoE. 





%



\bibliographystyle{IEEEtran}
\bibliography{IEEEabrv,tzeBib}

\end{document}